\documentclass[prd,aps, tightenlines, preprintnumbers, showpacs, nofootinbib,superscriptaddress,notitlepage]{revtex4-1}
  \usepackage{latexsym,bm,amsmath,amssymb,amsfonts}
\newcommand{\beq}{\begin{eqnarray}}
\newcommand{\eeq}{\end{eqnarray}}

\newcommand{\nn}{\nonumber \\}

\usepackage{graphicx}
\usepackage{amsmath,amsthm,amssymb}

\begin{document}
\preprint{YITP-14-37}

\title{Building up the elliptic flow: analytical insights}
\author{Yoshitaka Hatta}
\affiliation{Yukawa Institute for Theoretical Physics, Kyoto University, Kyoto 606-8502, Japan}

\author{Bo-Wen Xiao}
\affiliation{Key Laboratory of Quark and Lepton Physics (MOE) and Institute
of Particle Physics, Central China Normal University, Wuhan 430079, China}

\date{\today}
\vspace{0.5in}
\begin{abstract}
In this paper, we present a fully analytical description of the early-stage formation of elliptic flow in relativistic viscous hydrodynamics.
We first construct an elliptic deformation of Gubser flow which is a boost invariant solution of the Navier-Stokes equation with a nontrivial transverse profile. We then analytically calculate the momentum anisotropy of the flow as a function of time and discuss the connection with the empirical formula by Bhalerao {\it et al.} regarding the viscosity dependence of elliptic flow.
\end{abstract}
\maketitle

\section{Introduction}


One of the most intriguing results of ultrarelativistic heavy ion collisions at RHIC and the LHC is the strong collectivity of the created hot and dense matter, especially the considerable elliptic flow \cite{Ackermann:2000tr, Adcox:2002ms, Back:2002gz, Aamodt:2010pa, ATLAS:2011ah, Chatrchyan:2012wg}.
In non-central collisions, the overlapping region of the colliding nuclei approximately has the shape of an ellipse in the transverse plane.
This region  expands hydrodynamically, and the initial anisotropy in the pressure gradient gets converted into momentum space anisotropy, resulting in the modulation of the azimuthal angle distribution of charged particles in the final state \cite{Ollitrault:1992bk}
\begin{equation}
\frac{dN}{d\phi} \propto 1+2v_2 \cos 2\phi\,.
\end{equation}
The coefficient $v_2$ is called the elliptic flow parameter and is one of the central objects of experimental and theoretical study in heavy-ion physics because it is a sensitive probe of the equilibrium/nonequilibrium properties of the created matter.

While the mechanism to generate $v_2$ is  well understood, little is known about its  analytical details. Theoretically, the extraction of $v_2$ relies heavily on numerical (viscous) hydrodynamic simulations supplemented with some initial condition and the equation of state (see, e.g., \cite{Kolb:1999it,Kolb:1999es,Huovinen:2001cy,Teaney:2003kp,Gombeaud:2007ub,Xu:2007jv,
Luzum:2008cw,Song:2008hj,Hirano:2010jg,Niemi:2011ix}). In this paper, we provide a fully analytical description of the early-stage formation of $v_2$  by deriving and utilizing an approximate elliptic solution of the relativistic Navier-Stokes equation. Such an analysis has long been infeasible due to the difficulty of constructing realistic elliptically-shaped solutions of hydrodynamic equations for which only a few attempts have been made in the literature \cite{Csorgo:2003ry, Csanad:2003qa, Sin:2006pv,Peschanski:2009tg}. The new solution we are able to present here has been achieved along the line of the recent progress in constructing exact solutions of viscous hydrodynamic equations using conformal symmetry \cite{Gubser:2010ze,Gubser:2010ui,Marrochio:2013wla,Hatta:2014gqa,Hatta:2014gga}.
 Within the region of validity of the solution, one can study explicitly how the elliptic flow develops as a function of proper time and how the shear viscosity affects this evolution.

  In Section II, we review the Gubser flow \cite{Gubser:2010ze,Gubser:2010ui} which is an exact boost-invariant solution of the relativistic Navier-Stokes equation with nontrivial radial flow velocities.
  In Section III, we use the so-called Zhukovsky transform to elliptically deform the Gubser flow in the transverse plane and construct an approximate solution. We then calculate in Section IV the momentum space anisotropy of this flow and discuss the connection to the empirical formula proposed by Bhalerao \emph{et al.} \cite{Bhalerao:2005mm} regarding the shear viscosity dependence of the elliptic flow. In the end, we summarise in Section V.

\section{Gubser flow}
In this section, we briefly review the exact boost-invariant  solution of the relativistic Navier-Stokes equation found by Gubser \cite{Gubser:2010ze,Gubser:2010ui}. The solution is naturally explained by rewriting the Minkowski metric as
\begin{equation}
d\hat{s}^2=-d\hat{t}^2+d\hat{x}^2+d\hat{y}^2+d\hat{z}^2= -d\hat{\tau}^2+
d\hat{x}_\perp^2 + \hat{x}_\perp^2 d\hat{\phi}^2 + \hat{\tau}^2d\hat{\zeta}^2\,,
\end{equation}
 where $\hat{\tau}\equiv \sqrt{\hat{t}^2-\hat{z}^2}$ is the proper time and $\hat{\zeta}\equiv\tanh^{-1}\hat{z}/\hat{t}$ is the space-time rapidity.
 In this coordinate system, the four-velocity $\hat{u}^\mu$ of the fluid (normalized as $\hat{u}_\mu \hat{u}^\mu=-1$) reads
\begin{equation}
\hat{u}_\tau=-\cosh\left[\tanh^{-1}\frac{2\hat{\tau}\hat{x}_\perp}{L^2+\hat{\tau}^2
+\hat{x}_\perp^2}\right]\,,\quad  \hat{u}_\perp=\sinh\left[\tanh^{-1} \frac{2\hat{\tau}\hat{x}_\perp}{L^2+\hat{\tau}^2+\hat{x}_\perp^2}\right]\,, \label{boo}
\end{equation}
 with $\hat{u}_\phi=\hat{u}_\zeta=0$ and $L$ is roughly the initial transverse size of the fluid. The  $\hat{\zeta}$-independence of Eq.~(\ref{boo}) and $\hat{u}_\zeta=0$ mean that the flow expands in the longitudinal ($\hat{z}$) direction in a boost-invariant way. It also expands in the  transverse direction with the transverse velocity
\begin{equation}
\hat{v}_\perp=-\frac{\hat{u}_\perp}{\hat{u}_\tau} = \frac{2\hat{\tau}\hat{x}_\perp}{L^2+\hat{\tau}^2+\hat{x}_\perp^2}\,.
\end{equation}
Plugging this velocity profile into the relativistic Navier-Stokes equation and assuming conformal symmetry, Gubser obtained the following exact solution for the energy density
\begin{equation}
\hat{{\mathcal E}}=\frac{1}{\hat{\tau}^4}\frac{C^4}{(\cosh\rho)^{8/3}}\left[1+\frac{\eta_0}{9C}(\sinh\rho)^3 \,_2F_1\left(\frac{3}{2},\frac{7}{6},\frac{5}{2};-\sinh^2\rho\right)\right]^4\,, \label{gub}
 \end{equation}
 where $C>0$ is a constant and the shear viscosity $\hat{\eta}$ has been made dimensionless by factoring out the corresponding power of the energy density $\eta_0\equiv \hat{\eta}/ \hat{{\mathcal E}}^{3/4}$. In Eq.~(\ref{gub}), we have defined
 \begin{equation}
 \sinh\rho \equiv-\frac{L^2-\hat{\tau}^2+\hat{x}_\perp^2}{2L\hat{\tau}}\,.
 \end{equation}
The following components of the shear tensor will be needed for a later calculation.
\beq
 \hat{\sigma}_{\tau\perp} &=& \frac{2}{3}\frac{\hat{x}_\perp(L^2-\hat{\tau}^2+\hat{x}_\perp^2)(L^2+\hat{\tau}^2+\hat{x}_\perp^2)}
{((L^2+\hat{\tau}^2-\hat{x}_\perp^2)^2+(2L\hat{x}_\perp)^2)^{3/2}}\,, \qquad \
\hat{\sigma}_{\perp\perp}=-\frac{1}{3\hat{\tau}}\frac{(L^2+\hat{\tau}^2+
\hat{x}_\perp^2)^2(L^2-\hat{\tau}^2+\hat{x}_\perp^2)}
{((L^2+\hat{\tau}^2-\hat{x}_\perp^2)^2+(2L\hat{x}_\perp)^2)^{3/2}}\,,
\nn && \qquad \quad \hat{\sigma}_{\perp \phi}=0\,, \qquad
 \hat{\sigma}_{\phi\phi}=-\frac{1}{3\hat{\tau}}\frac{\hat{x}_\perp^2 (L^2-\hat{\tau}^2+\hat{x}_\perp^2)}{\sqrt{(L^2+
\hat{\tau}^2-\hat{x}_\perp^2)^2+(2L\hat{x}_\perp)^2}}\,. \label{she}
\eeq

 As already pointed out by Gubser, the
 solution has a pathological behavior at large negative values of $\rho$ corresponding to large $\hat{x}_\perp$ and/or small $\hat{\tau}$. Indeed, when $|\sinh\rho|\gg 1$  one can approximate
\begin{equation}
\,_2F_1\left(\frac{3}{2},\frac{7}{6},\frac{5}{2};-\sinh^2\rho\right) \approx \frac{9}{2}(-\sinh \rho)^{-7/3} + {\mathcal O}(\sinh^{-3}\rho)\,,
\end{equation}
 so that
 \begin{equation}
 \hat{{\mathcal E}}\approx \frac{C^4}{\hat{\tau}^4 (\cosh^2\rho)^{4/3}}\left[1-\frac{\eta_0}{2C}\left\{(-\sinh\rho)^{2/3}+{\mathcal O}(1)\right\}\right]^4\,. \label{neg}
 \end{equation}
 The quantity inside the square brackets is proportional to the temperature $\hat{T}\propto \hat{{\mathcal E}}^{1/4}$ and this becomes negative for sufficiently negative values of $\rho$. This is actually not surprising since the relativistic (first-order) Navier-Stokes equation  is known to have unphysical features.\footnote{The problem of negative temperature also appears in an exact solution of the Navier-Stokes equation for the Bjorken flow \cite{Hatta:2014gga}. } As demonstrated in Ref.~\cite{Marrochio:2013wla}, the problem can be cured by switching to the (second-order) Israel-Stewart equation. For the present purpose, we are not concerned about this issue since one can consider $\eta_0$ to be arbitrarily small (or $C$ arbitrarily large) so that the temperature remains positive in a parametrically large region of $\hat{x}_\perp$ and $\hat{\tau}$.
 \begin{equation}
\frac{\hat{\tau} L}{L^2\   \mbox{or} \   \hat{x}^2_\perp} \gg \left(\frac{\eta_0}{C}\right)^{3/2}\,.
 \label{up}
 \end{equation}
 In this region, the solution is well-behaved and offers an attractive model for the studies of  strongly interacting matter created in heavy-ion collisions as discussed in Refs.~\cite{Gubser:2010ze,Gubser:2010ui}.

\section{Elliptic solution}

The Gubser solution described above is cylindrically symmetric around the $\hat{z}$-axis. Here we relax this restriction and construct an approximate solution which has the shape of an ellipse in the transverse plane. This can be achieved by employing the so-called Zhukovsky (Joukowski) transform\footnote{The Zhukovsky transform was originally used to determine the two-dimensional incompressible potential flow around an airfoil.} which maps a circle onto an ellipse as follows
\begin{equation}
\hat{x}=\left(x_\perp + \frac{a^2}{x_\perp}\right)\cos\phi = x+\frac{a^2x}{x^2+y^2}\,, \quad \hat{y}=\left(x_\perp -\frac{a^2}{x_\perp}\right)\sin\phi = y-\frac{a^2y}{x^2+y^2}\,, \label{zhu}
\end{equation}
 where $a$ is a constant and we only consider the region $x_\perp>a$.
  As is manifest in its complex representation $\hat{\omega}=\hat{x}+i\hat{y}=\omega+\frac{a^2}{\omega}$, Eq.~(\ref{zhu}) is a conformal transformation in two-dimensions and therefore the metric is preserved up to a Weyl factor
\begin{equation}
d\hat{x}^2+d\hat{y}^2= \left(1-\frac{2a^2}{x_\perp^2}\cos 2\phi +\frac{a^4}{x_\perp^4} \right)(dx^2+dy^2)\equiv
A^2 (dx^2+dy^2)\,.
\end{equation}
Embedding this in four-dimensions, we obtain
\beq
d\hat{s}^2&=&-d\hat{\tau}^2 + d\hat{x}^2 + d\hat{y}^2+\hat{\tau}^2d\hat{\zeta}^2 \nonumber \\
&=& A^2\left[ -d\tau^2+dx_\perp^2+x_\perp^2d\phi^2+\tau^2d\zeta^2 -\frac{2\tau}{A}d\tau dA -\frac{\tau^2}{A^2}(dA)^2 \right]\,, \label{mod}
\eeq
 where we have relabeled $\hat{\tau}=A\tau$ and $\hat{\zeta}=\zeta$. If the last two terms were absent, the metric inside the square brackets would be exactly Minkowskian. Let us find the conditions under which these terms can indeed be neglected. More explicitly, we find
\begin{equation}
dA= \frac{2a^2}{Ax_\perp^2} \left[\left(\cos 2\phi-\frac{a^2}{x_\perp^2}\right)\frac{dx_\perp}{x_\perp} +\sin 2\phi d\phi\right]\,.
\end{equation}
Since the elliptic deformation $\sim \cos2\phi$ is an ${\mathcal O}(a^2/x_\perp^2)$ effect, we assume $a^2/x_\perp^2 \ll 1$ and neglect terms of order ${\mathcal O}(a^4/x_\perp^4)$.
Then the $(dA)^2$ term in (\ref{mod}) can be safely dropped. In order to drop the cross term ${\mathcal O}(d\tau dA)$ as well, we must additionally assume that $x_\perp \gg \tau$ and neglect terms of order ${\mathcal O}(\tau a^2/x_\perp^3)$ relative to the leading term.

Under these conditions, the coordinate systems $(\hat{\tau},\hat{x},\hat{y},\hat{\zeta})$ and $(\tau,x,y,\zeta)$ are conformally related, and one can map the Gubser solution expressed in the former coordinates into the latter.\footnote{Since Eq.~(\ref{zhu}) is not an element of the M\"obius transformation, it cannot exactly be promoted to a four-dimensional conformal transformation. This is why the solution obtained is an approximate one.} 
The transformation rule is given by
\begin{equation}
  {\mathcal E}=A^4\hat{{\mathcal E}}\,,  \qquad u_\mu=\frac{1}{A}\frac{\partial \hat{x}^\nu}{\partial x^\mu}\hat{u}_\nu\,,  \qquad \sigma_{\mu\nu}=\frac{1}{A}\frac{\partial \hat{x}^\alpha}{\partial x^\mu}\frac{\partial \hat{x}^\beta}{\partial x^\nu} \hat{\sigma}_{\alpha\beta},.
\end{equation}
To the order of interest, we can approximate
\begin{equation}
\hat{\tau}=A\tau\approx \left(1-\frac{a^2}{x_\perp^2}\cos2\phi\right)\tau \,, \qquad \hat{x}_\perp \approx x_\perp+\frac{a^2}{x_\perp}\cos2\phi\,,
\end{equation}
\begin{equation}
 \frac{\partial \hat{\phi}}{\partial \phi}\approx
1-\frac{2a^2}{x_\perp^2} \cos 2\phi\,, \qquad  \frac{\partial \hat{\phi}}{\partial x_\perp}\approx \frac{2a^2}{x_\perp^3}\sin 2\phi\,,
\end{equation}
 \begin{equation}
 \sinh\rho \approx -\frac{L^2-\tau^2+x_\perp^2}{2L\tau}\left(1+
\frac{a^2(L^2+3x_\perp^2)}{x_\perp^2 (L^2+x_\perp^2)}\cos 2\phi \right)\,.
 \end{equation}
We thus find the energy density
\beq
 {\mathcal E}
&\approx&\frac{C^4}{\tau^{4/3}} \frac{(2L)^{8/3}}{(L^4+2(\tau^2+x_\perp^2)L^2+(\tau^2-x_\perp^2)^2
 )^{4/3}} \left(1-\frac{\eta_0}{2C} \left(\frac{L^2-\tau^2+x_\perp^2}{2L\tau}\right)^{2/3}\right)^4\nonumber \\
&&  \qquad \qquad \times \left(1-\frac{a^2}{1-\frac{\eta_0}{2C}\left(\frac{L^2-\tau^2+x_\perp^2}{2L\tau}\right)^{2/3}}
\frac{8}{3x_\perp^2}\frac{L^2+3x_\perp^2}{L^2+x_\perp^2}\cos2\phi  \right) \nonumber \\
&\equiv& {\mathcal E}_0+ \delta {\mathcal E} a^2\cos2\phi \,. \label{en}
\eeq
 Since $\delta {\mathcal E}<0$, the curves of constant energy are elliptic, with the major axis in the $y$-direction.
The flow velocity in turn is given by
\beq
 && u_\tau \approx -1+{\mathcal O}(\tau^2)\,, \nn
&&u_\perp
\approx
\frac{2\tau x_\perp}{L^2+\tau^2+x_\perp^2} -2\tau \left(\frac{1}{x_\perp^3}+\frac{2x_\perp}{(L^2+x_\perp^2)^2}\right)a^2\cos2\phi  \equiv u_{\perp 0}+\delta u_\perp a^2\cos2\phi \,, \nn
&& u_\phi \approx \frac{-2\tau }{x_\perp^2} \frac{L^2+3x_\perp^2}{L^2+x_\perp^2}a^2\sin 2\phi  \equiv \delta u_\phi a^2\sin2\phi \,,
\label{flow}
\eeq
 and the shear tensor is
\beq
\sigma_{\perp\perp}
&\approx& -\frac{1}{3\tau}
\left(1-\frac{2(L^2-2x_\perp^2)\tau^2}{(L^2+x_\perp^2)^2} -
\frac{4\tau^2(L^4+L^2x_\perp^2+6x_\perp^4)}{(L^2+x_\perp^2)^3 x_\perp^2} a^2 \cos2\phi\right) \equiv \sigma_{\perp\perp}^0+\delta \sigma_{\perp\perp} a^2\cos2\phi \,,\nn
\sigma_{\perp \phi}
&\approx& \frac{4 \tau (L^2+3x_\perp^2)}{3x_\perp(L^2+x_\perp^2)^2}a^2 \sin 2\phi \equiv \delta \sigma_{\perp \phi} a^2\sin2\phi \,, \nn
 \sigma_{\phi\phi}
&\approx& -\frac{x_\perp^2}{3\tau}\left(1-\frac{2L^2\tau^2}{(L^2+x^2_\perp)^2} +\frac{4\tau^2 L^2(L^2+3x_\perp^2)}{(L^2+x_\perp^2)^3 x_\perp^2} a^2 \cos 2\phi\right) \equiv \sigma_{\phi\phi}^0+ \delta \sigma_{\phi\phi} a^2\cos2\phi\,. \label{st}
\eeq

Note that $\delta u_\perp$ is negative, meaning that both ${\mathcal E}$ and $u_\perp$ are stretched in the $y$-direction.
This may seem contradictory to the standard picture that the stronger pressure (or energy density) gradient in the $x$-direction develops a stronger flow in the same direction. In fact, there is no contradiction. The negativeness of $\delta u_\perp$ (at large $x_\perp$) is a direct consequence of conformal symmetry\footnote{The situation is different in the confining (non-conformal) case. Suppose that the energy density decays as a Gaussian instead of a power-law
\beq
{\mathcal E} \sim e^{-\frac{x^2}{\sigma_x^2}-\frac{y^2}{\sigma_y^2}}\approx \exp\left(-\frac{x_\perp^2}{2}\left(\frac{1}{\sigma_x^2}+\frac{1}{\sigma_y^2}\right) \right)
\left(1-\frac{x_\perp^2}{2}\left(\frac{1}{\sigma_x^2}-\frac{1}{\sigma_y^2}\right)\cos 2\phi \,, \nonumber \right)\,,
\eeq
 where $\sigma_y > \sigma_x$. The Euler equation is then
 \beq
 u_\perp \sim -\tau \partial_\perp \ln  {\mathcal E}\sim \tau x_\perp \left(\frac{1}{\sigma_x^2}-\frac{1}{\sigma_y^2}\right)\cos 2\phi =\delta u_\perp \cos 2\phi\,, \nonumber
 \eeq
  for the $\phi$-dependent part. 
 Thus $\delta u_\perp$  is positive in this case.  \label{footnote}
} which dictates a power-law decay of ${\mathcal E}$ and the hydrodynamic equation which, in the present accuracy, boils down to
\begin{equation}
\delta u_\perp = -\frac{3\tau}{8}\partial_\perp  \left(\frac{\delta{\mathcal E}}{{\mathcal E}_0}\right)  <0\,. \label{equ}
\end{equation}
(For simplicity here we consider the ideal hydrodynamic equation.)
On the other hand, we find
\beq
\int_0^{2\pi} d\phi \,(u_x^2-u_y^2) &\approx&\int d\phi \left(\cos2\phi \,u_\perp^2-\frac{2\sin2\phi}{x_\perp}u_\phi u_\perp \right) \nn &=&
2\pi a^2 u_{\perp 0}\left(\delta u_\perp -\frac{\delta  u_{\phi}}{x_\perp}\right)= \frac{16\pi a^2\tau^2 L^2}{(L^2+x_\perp^2)^3}>0\,, \label{ux}
\eeq
 where we have neglected $u_\phi^2 \sim {\mathcal O}(a^4)$.
 Eq.~(\ref{ux}) shows that, due to the $\phi$-component of the flow velocity which tends to point to the $x$-direction ($\delta u_\phi<0$), the flow is on average stronger in the $x$-direction, as expected.

Eqs.~(\ref{en})-(\ref{st}) define an approximate elliptic solution of the Navier-Stokes equation with the shear viscosity $\eta=\eta_0 {\mathcal E}^{3/4}$. They satisfy the equation with the accuracy of order ${\mathcal O}(a^2/x_\perp^2)$ relative to the leading (isotropic) solution, but break down at order ${\mathcal O}(a^2\tau^2/x_\perp^4)$ which can be checked explicitly.\footnote{Note that we have already neglected terms of order ${\mathcal O}(\tau^2/x_\perp^2)$ in the ${\mathcal O}(a^2)$ corrections in Eqs.~(\ref{en})-(\ref{st}).} One way to understand this is to notice that there is in fact an ${\mathcal O}(\tau^2/x_\perp^2)$ uncertainty in the definition of $a^2$. Indeed, we could have employed a nonconstant  $a^2(\tau)$ already in Eq.~(\ref{zhu}) as long as its $\tau$-dependence is sufficiently weak
 \begin{equation}
 \frac{1}{a^2}\frac{\partial a^2}{\partial\tau} \lesssim {\mathcal O}\left(\frac{\tau}{x_\perp^2}\right)\,. \label{con}
 \end{equation}
If Eq.~(\ref{con}) is satisfied, the extra terms that would appear in the transformation as in Eq.~(\ref{mod}) are of the same order as, or less than those already neglected. The maximal uncertainty incurred by Eq.~(\ref{con}) is
\begin{equation}
a^2(\tau)\sim a_0^2\left( 1+{\mathcal O}\left(\frac{\tau^2}{x_\perp^2}\right) \right)\,, \label{quad}
\end{equation}
and this affects the Navier-Stokes equation at order ${\mathcal O}(a^2\tau^2/x_\perp^4)$.

A similar caveat applies to the eccentricity of the flow which we define as
 \begin{equation}
\epsilon(\tau) \equiv \frac{\langle y^2-x^2\rangle}{\langle y^2+x^2\rangle} \equiv \frac{\int dxdy (y^2-x^2){\mathcal E}(x_\perp,\tau)}{\int dx dy (y^2+x^2) {\mathcal E}(x_\perp,\tau)}\,.
\label{ec}
\end{equation}
For the ideal solution at early times, we find
\begin{equation}
\epsilon^{ideal}(\tau \approx 0) \approx \frac{14a^2}{9L^2}\,.
\end{equation}
 On general grounds, one expects that $\epsilon(\tau)$ decreases with time. Unfortunately, this $\tau$-dependence is not reliably calculable in the present approach due to the above uncertainty in $a^2$. We however note that Eq.~(\ref{quad}) suggests that $\epsilon(\tau)$ decreases quadratically in $\tau$
  \begin{equation}
  \epsilon(\tau) \sim \epsilon_0 \left(1-c\frac{\tau^2}{L^2}\right)\,,
 \label{dec} \end{equation}
  where $c>0$ remains undetermined. Eq.~(\ref{dec}) appears to be consistent with the result of numerical simulations \cite{Kolb:1999es,Luzum:2008cw} (see also, \cite{Taliotis:2010pi}).

Before proceeding to the calculation of elliptic flow with viscous corrections, we  make one more simplification.
In order to clearly see the role of the viscosity, we shall neglect the terms of order ${\mathcal O}(\tau^2/x_\perp^2)$ and ${\mathcal O}(\tau^2/L^2)$ in the energy density Eq.~(\ref{en})  while keeping powers of  $\eta_0 (L/\tau)^{2/3}$ up to quadratic order. Actually, once we go beyond linear order in $\eta_0$, we must be careful with the ${\mathcal O}(1)$ terms inside the brackets of Eq.~(\ref{neg}). The condition that these terms are parametrically smaller than the ${\mathcal O}(\eta^2_0(L/\tau)^{4/3})$ term is
\begin{equation}
\left(\frac{\eta_0}{C}\right)^{3/4} \gg \frac{\tau}{L\ \mbox{or}\ x_\perp} \gg \left(\frac{\eta_0}{C}\right)^{3/2}\,, \label{wi}
\end{equation}
where another constraint in Eq.~(\ref{up}) has been included. Thus, strictly speaking, the following analysis is valid in the window as shown in Eq.~(\ref{wi}) in the viscous case with $\eta_0 \neq 0$.

\section{Viscous effect on elliptic flow}

We now turn to the discussion of the elliptic flow parameter $v_2$ which characterizes the momentum space anisotropy in the final state. Besides the leading contribution from the ideal flow, we also include the shear viscous correction which plays an important role in the phenomenological study of quark gluon plasma \cite{Teaney:2003kp, Xu:2007jv, Luzum:2008cw,Song:2008hj, Niemi:2011ix}.

With an analytic solution of the Navier-Stokes equation at hand, we can get information about $v_2$ via a closely related quantity  \cite{Kolb:1999it,Kolb:1999es}
\begin{equation}
\epsilon_p(\tau) \equiv \frac{\int dx dy (T_{xx}-T_{yy}) }{\int dx dy ( T_{xx}+T_{yy})}\,. \label{v2}
\end{equation}
Explicitly, we find
\beq
T_{xx}+T_{yy}&\approx& \frac{2{\mathcal E}}{3}\left(1+2u_\perp^2\right)+\pi_{\perp\perp}+\frac{1}{x_\perp^2}\pi_{\phi\phi}\,,  \label{den}\\
T_{xx}-T_{yy} &=& \frac{4{\mathcal E}}{3}(u_x^2 - u_y^2) + \pi_{xx}-\pi_{yy} \nn
 &\approx& \cos2\phi \left(\frac{4{\mathcal E}}{3}u_\perp^2 + \pi_{\perp\perp}-\frac{1}{x_\perp^2}\pi_{\phi\phi} \right)-\frac{2\sin2\phi}{x_\perp}  \left(\frac{4{\mathcal E}}{3}u_\perp u_\phi  +\pi_{\perp \phi}\right)\,, \label{num}
\eeq
 where  $\pi^{\mu\nu}$ is the shear-stress tensor.
Unlike $v_2$, $\epsilon_p(\tau)$ in Eq.~(\ref{v2}) is defined at all times, and its value at the `build-up time'
 \begin{equation}
 \tau_f \sim \frac{L}{c_s}\,, \label{build}
\end{equation}
where $L$ represents the characteristic transverse size and $c_s$ is the speed of sound, is considered to be a measure of $v_2 \propto \epsilon_p(\tau_f)$.  ($c_s=1/\sqrt{3}$ in a conformal theory.)
However, as already mentioned, $\tau_f\sim {\mathcal O}(L)$ is outside the region of validity for our approximate solution. Nevertheless, we show that Eq.~(\ref{v2}) evaluated at $\tau\ll \tau_f$ provides us with useful analytical insights into the property of $\epsilon_p(\tau)$ as we extrapolate $\tau\to \tau_f$.

First, let us consider the denominator in Eq.~(\ref{den}). Since $\epsilon_p\propto a^2$ and the numerator is ${\mathcal O}(a^2)$, we can set $a=0$ in the denominator from the beginning. We then recall that in the Navier-Stokes approximation,
 \beq
& \pi_{\mu\nu}=-2\eta \sigma_{\mu\nu}
=-2\eta_0 {\mathcal E}^{3/4} \sigma_{\mu\nu}\,,
 \eeq
  where $\sigma_{\mu\nu}$ is given by Eq.~(\ref{st}) and
\beq
 {\mathcal E}^{3/4}
&\approx&\frac{C^3}{\tau} \frac{(2L)^{2}}{(L^2+x_\perp^2)^{2}} \left(1-\frac{\eta_0}{2C} \left(\frac{L^2+x_\perp^2}{2L\tau}\right)^{2/3}\right)^3\nonumber \\
&&  \qquad \qquad \times \left(1-\frac{a^2}{1-\frac{\eta_0}{2C}\left(\frac{L^2+x_\perp^2}{2L\tau}\right)^{2/3}}
\frac{2}{x_\perp^2}\frac{L^2+3x_\perp^2}{L^2+x_\perp^2}\cos2\phi  \right) \nonumber \\
&\equiv& {\mathcal E}_0^{3/4}+ \delta {\mathcal E}^{3/4} a^2\cos2\phi \,, \label{en2}
\eeq
 in the present approximation.
  Moreover, $u_\perp^2 \sim {\mathcal O}(\tau^2)$ in the factor $1+2u_\perp^2$  can be neglected. It is then straightforward to show that
  \begin{equation}
\int dxdy (T_{xx}+T_{yy})  \approx \frac{8\pi C^4}{5\tau^{4/3}}\left(\frac{2}{L}\right)^{2/3}
\left(1-\frac{15\eta_0^2}{2C^2}\left(\frac{L}{2\tau}\right)^{4/3}\right) \,.
\label{tt}
\end{equation}
Note that the viscous correction shows up only at ${\mathcal O}(\eta_0^2)$.\footnote{In fact, the ${\mathcal O}(\eta_0^3)$ term of the denominator (but not the numerator) is divergent because the  $x_\perp$-integral does not converge as $x_\perp \to \infty$. This is an artifact of the solution Eq.~(\ref{neg}) which becomes unphysical at large $x_\perp$, and limits our calculation of $\epsilon_p$  to
${\mathcal O}(\eta_0^2)$. }

We now turn to the numerator which requires some extra care. After the $\phi$ integral, we get
\beq
&& \int d\phi (T_{xx}-T_{yy}) = \frac{4\pi a^2 u_{\perp 0}}{3}\left\{ \delta {\mathcal E} u_{\perp0}
+2{\mathcal E}_0 \left(\delta u_\perp -\frac{1}{x_\perp} \delta u_\phi\right) \right\} \nn
&& \qquad \qquad \qquad \qquad  -2\pi a^2 \eta_0 \left\{ \left(\sigma_{\perp\perp}^0 -\frac{\sigma_{\phi\phi}^0}{x_\perp^2}\right)
 \delta {\mathcal E}^{3/4} + \left(\delta \sigma_{\perp\perp} -\frac{\delta \sigma_{\phi\phi}}{x_\perp^2} -\frac{2 \delta \sigma_{\perp \phi}}{x_\perp} \right) {\mathcal E}_0^{3/4}  \right\} \nn
 && \qquad = \frac{16\pi a^2\tau^2}{3(L^2+x_\perp^2)^2}\left(x_\perp^2 \delta {\mathcal E} + \frac{4L^2 {\mathcal E}_0}{L^2+x_\perp^2}\right)+\frac{8\pi a^2 \eta_0 \tau}{3(L^2+x_\perp^2)^2} \left(x_\perp^2 \delta {\mathcal E}^{3/4} + \frac{4L^2 {\mathcal E}_0^{3/4} }{L^2+x_\perp^2} \right)\,, \label{reco}
\eeq
where the potentially dangerous ${\mathcal O}(1/x_\perp^2)$ terms in Eqs.~(\ref{flow}) and (\ref{st}) which could cause trouble in the remaining $dx_\perp^2$ integral have canceled out. In Eq.~(\ref{reco}), we recognize two types of contributions with clear but distinct physical interpretations. The terms proportional to $\delta {\mathcal E}$ and $\delta {\mathcal E}^{3/4}$ are due to the spatial anisotropy of the source (energy density). Since the bulk of matter is initially stretched in the $y$-direction ($\delta {\mathcal E}<0$), these terms give a negative contribution to $\epsilon_p$. On the other hand, the terms proportional to ${\mathcal E}_0$ and ${\mathcal E}_0^{3/4}$ are due to the anisotropy of the flow velocity $\delta u$. As already discussed in Eq.~(\ref{ux}), they give a positive contribution to $\epsilon_p$.

We thus evaluate the two contributions separately and find, after dividing by Eq.~(\ref{tt}),
\begin{equation}
\left.\frac{\int dx dy (T_{xx}-T_{yy}) }{\int dx dy ( T_{xx}+T_{yy})}\right|_{\delta {\mathcal E}} = \frac{20 a^2\tau^2}{3L^4}
\left[ -\frac{80}{77}+ \frac{3\eta_0}{2C}\left(\frac{L}{2\tau}\right)^{2/3} -\frac{3264\eta_0^2}{385C^2}\left(\frac{L}{2\tau}\right)^{4/3} \right]\,, \label{deltae} 
\end{equation}
and
\begin{equation}
\left.\frac{\int dx dy (T_{xx}-T_{yy}) }{\int dx dy ( T_{xx}+T_{yy})}\right|_{\delta u}
=\frac{20 a^2\tau^2}{3L^4}
\left[ \frac{6}{7}- \frac{3\eta_0}{2C}\left(\frac{L}{2\tau}\right)^{2/3} +\frac{513\eta_0^2}{70C^2}\left(\frac{L}{2\tau}\right)^{4/3} \right]\,. \label{deltau} 
\end{equation}
Summing the two contributions, we arrive at
\begin{equation}
\epsilon_p(\tau)= \frac{20 a^2\tau^2}{3L^4}\left[-\frac{2}{11} -\frac{177\eta_0^2}{154C^2}\left(\frac{L}{2\tau}\right)^{4/3}\right]\,. \label{fin}
\end{equation}
Surprisingly, $\epsilon_p$ is negative as a result of the slightly larger contribution from the source anisotropy.  This is at odds with the observed behavior  in hydrodynamic simulations \cite{Kolb:1999it,Gombeaud:2007ub,Luzum:2008cw} which is  qualitatively quite consistent with the contribution from the flow anisotropy (\ref{deltau}) \emph{alone}, namely, $\epsilon_p(\tau)$ is positive, grows quadratically in time $\epsilon_p(\tau)\sim \tau^2$, and the viscosity tends to suppress it.\footnote{
Readers may wonder whether the results in Eqs.~(\ref{deltae})-(\ref{fin}), which are of the order of ${\mathcal O}(a^2\tau^2/L^4)$, are reliable in view of the difficulty we have encountered at this order when analyzing the hydrodynamic equation and the eccentricity. However, in the previous examples, the uncertainty at order ${\mathcal O}(a^2\tau^2/L^4)$ stems from  unknown corrections to the lower order ${\mathcal O}(a^2/L^2)$ results. In contrast, Eqs.~(\ref{deltae})--(\ref{fin}) are the leading order results for the momentum anisotropy. The would-be lower order terms of order ${\mathcal O}(\tau^2/L^2)$ and ${\mathcal O}(a^2/L^2)$, as well as their uncertainty have been canceled. It is also straightforward to see that the genuine ${\mathcal O}(a^2\tau^2/L^4)$ corrections of $\mathcal{E}$ and $u^\mu$ can only enter at even higher order in the results in Eqs.~(\ref{deltae})-(\ref{fin}).  }

 While this discrepancy may seem worrisome, one should notice the large cancelation which resulted  in a barely negative value found in (\ref{fin}). This suggests that  whether the source anisotropy contribution is large enough to flip the sign of $\epsilon_p$ is subtle and   model-dependent. It is then interesting that our conformal solution reveals and exemplifies the logical possibility that even if $\int d^2x_\perp (u_x^2-u_y^2)$ is positive, $\int d^2x_\perp {\mathcal E}(u_x^2-u_y^2)$ can become negative. When this occurs, the simple proportionality between $v_2$ and $\epsilon_p$ is far from obvious and may be subject to large `non-flow' effects (see e.g., Refs.~\cite{Hirano:2010jg, Huovinen:2001wn}).


What are, then,  the main characteristics of a given model which determine the sign of $\epsilon_p$? In the model at hand, the transverse flow $u_x^2-u_y^2$ generated by the pressure gradient is weak and insufficient to make $\epsilon_p$ positive because  $\delta u_\perp$ is negative  (cf., (\ref{ux})). This is an artifact of conformal symmetry which dictates that the energy density should decay as a power-law. In confining theories including QCD, $\delta u_\perp$ will be positive (see Footnote \ref{footnote}), and one therefore expects that the flow anisotropy contribution (\ref{deltau}) dominates over the source anisotropy contribution (\ref{deltae}).
 After all, elliptic flow is the anisotropy in the flow velocity, and this is faithfully incorporated in (\ref{deltau}), but not in (\ref{deltae}).\footnote{Another  difference from hydrodynamic simulations is  the treatment of the initial velocity.
  In typical numerical simulations, the transverse velocity at the initial time $\tau=\tau_0$ is set to be zero,  whereas in our solution the velocity (\ref{flow}) is nonzero for any $\tau_0>0$. Simulations with nonvanishing $u_\perp (\tau_0)$ do exist (see, e.g., \cite{Qin:2010pf,Pang:2012he,Gale:2012rq}), but so far they only studied $v_2$ in the final state. It would be interesting to see the effect of the initial velocity on the early-stage development of $\epsilon_p$ in these simulations.}

 Let us therefore take a closer look at Eq.~(\ref{deltau}). Parametrically, and to linear order in $\eta_0$, this can be rewritten as
\begin{equation}
\left.\frac{\epsilon_p(\tau)}{\epsilon_p^{ideal}(\tau)} \right|_{\delta u}\sim \frac{1}{1+\frac{\eta_0}{C} \left(\frac{L}{\tau}\right)^{2/3}}
\sim \frac{1}{1+\frac{\eta L^2}{C^3{\mathcal E}^{1/4}} } \sim \frac{1}{1+\frac{ L^2}{\sigma dN/dY }}\,, \label{nice}
\end{equation}
 where  we have used ${\mathcal E}\sim T^4\sim C^4/(\tau^{4/3}L^{8/3})$ and the kinetic theory relation $\eta/{\mathcal E}^{1/4} \sim \eta/T \sim 1/\sigma$ (with $\sigma$ being the cross section of microscopic degrees of freedom) together with an estimate $C^3 \sim dN/dY$ of the rapidity distribution of particle multiplicity in the final state (see, Eq.~(44) of Ref.~\cite{Gubser:2010ze}).
Now let us compare Eq.~(\ref{nice}) with the following empirical formula proposed in heavy-ion collisions \cite{Bhalerao:2005mm} (and also in high-multiplicity proton-proton collisions \cite{Avsar:2010rf})
\begin{equation}
\frac{v_2}{v_2^{ideal}}=\frac{1}{1+ \frac{K}{K_0}}\,,\qquad K\equiv  \frac{S_\perp}{c_s \sigma \frac{dN}{dY}}\,,   \label{emp}
\end{equation}
where $K_0>0$ is a number of order unity and $S_\perp$ is the initial transverse overlap area of the two colliding nuclei. Despite its simplicity, the formula in Eq.~(\ref{emp}) is quite successful in fitting the RHIC/LHC data and the results of viscous hydrodynamic simulations \cite{Bhalerao:2005mm,Gombeaud:2007ub,Drescher:2007cd}.  With the natural identification $S_\perp=L^2$, parametrically the agreement between Eqs.~(\ref{nice}) and (\ref{emp}) is perfect.\footnote{The factor of $c_s$ in (\ref{emp}) originates from the build-up time Eq.~(\ref{build}) of the elliptic flow. Our solution, valid for $\tau\ll \tau_f$, does not have the accuracy to say anything about this ${\mathcal O}(1)$ factor. } Thus we have presented an explicit justification of the empirical formula Eq.~(\ref{emp}) to linear order in the `Knudsen' number  $K\sim \eta_0(L/\tau)^{2/3}$. At higher order in $K$, our result suggests a power series with alternating sign  (remember that ${\mathcal E}\sim (1-K)^4$ before the $d^2x_\perp$ integration). Since this is not a geometric series, we expect a deviation from the formula Eq.~(\ref{emp}) starting from ${\mathcal O}(K^2)$. Nevertheless, Eq.~(\ref{emp}) does capture the main feature of the series and this is probably the reason of its successful applications in phenomenology.

\section{Conclusion}
In conclusion, we have constructed an approximate boost-invariant elliptic solution of the Navier-Stokes equation with a well-defined region of validity. In this region,
the leading contribution to the momentum anisotropy (\ref{v2}) of the fluid is analytically calculated. Driven by the spatial anisotropy, a stronger flow develops in the direction of larger pressure gradient. While this scenario is well-known and well-tested numerically, in the presence of viscosity it has not been demonstrated  with the level of analytical precision  we have been able to present in this paper. We also pointed out the potentially large negative contribution to $\epsilon_p$ due to the source anisotropy which, depending on models, can even flip the sign of $\epsilon_p$. Finally, by focusing on the flow anisotropy contribution in Eq.~(\ref{deltau}), we have discussed the connection with the empirical formula (\ref{emp}) previously suggested in heavy-ion phenomenology. \\

\vspace{3mm}
\noindent \textbf{Acknowledgements}\\

We are grateful to Jorge Noronha and Jean-Yves Ollitrault for many helpful discussions. We also thank Tetsufumi Hirano and Guang-You Qin for discussions and comments.

\end{document}